\documentclass[runningheads]{PMS2026}

\usepackage[ansinew]{inputenc}
\usepackage[top=3cm,bottom=3.5cm,left=4cm,right=3.8cm,dvips]{geometry}
\usepackage{nopageno}

\usepackage{index}
\usepackage{fancyhdr}
\usepackage{indentfirst}
\usepackage{sectsty}
\allsectionsfont{\usefont{OT1}{phv}{bc}{n}\selectfont \normalsize \bf }
\usepackage{graphicx}
\usepackage{amsmath}
\usepackage{subcaption}
\usepackage{url}
\usepackage{amssymb}
\usepackage{epsfig}
\usepackage{hyperref}
\usepackage[agsmcite,abbr]{harvard}
\usepackage[capitalise]{cleveref} 
\usepackage{booktabs}

\newcommand{\taskSet}[0]{\mathcal{J}}
\newcommand{\taskPeriodSet}[1]{\mathcal{J}_{#1}}
\newcommand{\task}[1]{J_{#1}}
\newcommand{\taskPeriod}[1]{T_{\alpha_{#1}}}
\newcommand{\periodSet}[0]{\mathcal{T}}
\newcommand{\period}[1]{T_{#1}}
\newcommand{\proc}[1]{p_{i}}
\newcommand{\groupPeriodSet}[1]{\mathcal{G}_{#1}}
\newcommand{\group}[2]{G_{#1,#2}}
\newcommand{\groupSize}[2]{gs_{#1,#2}}
\newcommand{\tasksOfGroup}[2]{\mathcal{J}_{{#1},{#2}}}
\newcommand{\cmax}[0]{C_{\max}}
\newcommand{\header}[0]{hs}
\newcommand{\maxGroupSize}[0]{MG}
\newcommand{\binPeriodSet}[1]{\mathcal{B}_{#1}}
\newcommand{\binPeriodSize}[1]{|\binPeriodSet{#1}|}
\newcommand{\integers}[0]{\mathbb{Z}}
\newcommand{\binary}[0]{\left\{0,1\right\}}
\makeindex

\begin{document}
\pagenumbering{gobble}
\pagestyle{headings}
\mainmatter


\title{Periodic Scheduling of Grouped Time-Triggered Signals on a Single Resource}

\author{ Josef Grus\inst{1,2}, Zdenek Hanzalek\inst{2} and Claire Hanen\inst{3,4} }

\institute{DCE, FEE, Czech Technical University in Prague\\
\and
IID, CIIRC, Czech Technical University in Prague\\
\and
Sorbonne University, CNRS, LIP6\\
\and
University Paris Nanterre\\
\email{josef.grus@cvut.cz, zdenek.hanzalek@cvut.cz, claire.hanen@lip6.fr}
}

\maketitle
\index{Josef Grus}\index{Zdenek Hanzalek}\index{Claire Hanen}

\noindent
\textbf{Keywords:} bin packing, periodic scheduling, signal grouping

\section{Grouping of Time-Triggered Signals}

Time-triggered messages are of crucial importance in modern communication networks. Offline-generated schedules, which specify start times for periodic messages, enable us to achieve deterministic behavior in critical applications. In automotive and avionics domains, so-called signals (measurements and commands) are periodically generated and communicated (via messages) among sensors, controllers, and actuators. 

However, the message contains not only the useful signal data, but also necessary metadata, e.g., message ID. Metadata is stored as a header or tail and extends the message size; when the signal is very short (as it often is in applications), sending each in a separate message is inefficient. Thus, several signals are grouped into a single message, depending on their periodicity and length, and sent with just one header. Such an approach increases the utilization of the communication resource (link or bus), since less bandwidth is wasted on headers \cite{perf-an-grouping}. However, grouping the signals into messages is complicated. The maximum size of the message (including the metadata) is finite, since longer messages have a lower probability of successful delivery. Also, longer messages are less flexible for scheduling in a periodic setting. This is similar to the work of \citeasnoun{bundling-delay}, where the compromise between energy efficiency and latency for IoT devices was investigated.

In this paper, we study the fundamental problem of grouping time-triggered signals into messages and periodic scheduling of messages on a single resource.

\section{Problem Statement}
The joint problem of grouping signals and scheduling messages is modeled as an extension of the Periodic Scheduling Problem (PSP), which is generally strongly NP-complete \cite{cai1996nonpreemptive}. Our problem consists of set of tasks $\taskSet = \left\{\task{1},\dots,\task{i},\dots,\task{n}\right\}$. Each task models one of the time-triggered signals. $\task{i}$ is associated with integer period $\taskPeriod{i}$ and integer processing time $\proc{i}$. We consider a set of harmonic periods, which is often used in applications; for any $\task{i},\task{i'}$, $\taskPeriod{i}$ is an integer multiple of $\taskPeriod{i'}$, or vice versa. This implies period set $\periodSet=\left\{\period{0},\dots,\period{u},\ldots,\period{r-1}\right\}$ of size $r$, such that $\forall k\in\{1,\ldots,r-1\}, \period{u}=b_u\period{u-1}, b_u \in \mathbb{Z}^+$. $\alpha_i$ maps $\task{i}$ to one of the $r$ periods, partitioning $\taskSet$ to task-period sets $\taskPeriodSet{u}=\left\{\task{i}: \taskPeriod{i}=\period{u}\right\}$.

We allow grouping of signals with the same period only. Thus, tasks of $\taskPeriodSet{u}$ are grouped into groups (i.e., messages) given by set $\groupPeriodSet{u}=\left\{\group{u}{1},\dots\,\group{u}{j},\dots,\group{u}{|\taskPeriodSet{u}|}\right\}$; sufficient number of groups is used to accommodate single-task-per-group corner case. Group $\group{u}{j}$ has period $\period{u}$, and its processing time depends on tasks assigned to it. With a set of tasks $\tasksOfGroup{u}{j}$ assigned to $\group{u}{j}$, we define the group size $\groupSize{u}{j}$ as:

\begin{equation}
    \groupSize{u}{j} =
  \begin{cases}
    \header + \sum_{\task{i}\in\tasksOfGroup{u}{j}} \proc{i}  & |\tasksOfGroup{u}{j}| > 0 \\
    0 & \text{otherwise}
  \end{cases}
\end{equation}

Thus, $\groupSize{u}{j}$ includes the header of constant size $hs$, if there are any tasks, and otherwise is considered unused/empty. The group size is also upper-bounded by a fixed maximum group size $\maxGroupSize$. Every task is part of some group: $\taskPeriodSet{u}=\bigcup_{\forall\group{u}{j}}\tasksOfGroup{u}{j}$. The groups are then scheduled on a resource. Our grouping differs from grouping in the context of a multi-processor scheduling \cite{grouping-to-multiproc}, where task sets are partitioned among multiple resources. As in \citeasnoun {grus2024icores} the schedule can be visualized by stacking each of the first $\frac{\period{r-1}}{\period{0}}$ intervals of length $\period{0}$ on top of another. A row $k$  is then called an observation interval and corresponds to the interval $[k-1\cdot\period{0},k\cdot\period{0})$. Each group of period $T_u$  is scheduled periodically in $\frac{\period{r-1}}{\period{u}}$ such intervals. The shorter the period of the group is, the more occurrences of it are present. This is shown in \cref{fig:example}


As was applied in \citeasnoun{lukasiewycz2009flexray}, \citeasnoun{eisenbrand2010solving}, and proven in \citeasnoun{grus2024icores}, only solutions in canonical order need to be considered. Canonical order means that each task/group of a given period has all the tasks/groups with a shorter period to its left in each observation interval. Thus, scheduling groups into suitable observation intervals can be modeled similarly to bin packing. It is only necessary to decide in which observation interval the first occurrence of the group is scheduled (the others are scheduled implicitly). The indices of intervals relevant for the first occurrence are given by the set $\binPeriodSet{u}=\left\{0,\dots,\period{u}/\period{0} - 1\right\}$.

To find a feasible solution, each observation interval utilization (sum of processing time of occurrences of present groups) needs to be limited by the shortest period $\period{0}$. However, this constraint can be made into a soft constraint by minimizing the $\cmax$, the maximum utilization of any observation interval. If $\cmax \le \period{0}$, the solution is feasible with respect to the minimum period. Any margin $\period{0}-\cmax$ makes the schedule more flexible, e.g., for scheduling additional sporadic tasks.

\section{Mathematical Programming Model}
In this section, we describe the mixed-integer linear programming (MILP) model for the extended PSP outlined in the previous section. The model assigns task $\task{i}$ to exactly one group $\group{u}{j}$ via binary variables  $x_{uij}$ in \cref{taskIsScheduled}. Group size $\groupSize{u}{j}$ is calculated and constrained with \cref{signalVar,groupSizeDefined,groupSizeDefined2}, where the header is included via binary variable $z_{uj}$.

\vspace{-10pt}
{\small
\begingroup
\allowdisplaybreaks
\begin{align}
    &\min \cmax\label{obj}\\
    & \sum_{\forall  \group{u}{j} \in \groupPeriodSet{u}} x_{uij} = 1 & \forall \period{u} \in \periodSet~\forall\task{i}\in\taskPeriodSet{u} \label{taskIsScheduled}\\
    & \sum_{\forall\task{i}\in\taskPeriodSet{u}} \proc{i}\cdot x_{uij} \le |\taskPeriodSet{u}| \cdot z_{uj} &\forall \period{u} \in \periodSet~\forall\group{u}{j}\in\groupPeriodSet{u}\label{signalVar}\\
    & \groupSize{u}{j} = \header \cdot z_{uj} + \sum_{\forall\task{i}\in\taskPeriodSet{u}} \proc{i}\cdot x_{uij} &\forall \period{u} \in \periodSet~\forall\group{u}{j}\in\groupPeriodSet{u}\label{groupSizeDefined}\\
    & \groupSize{u}{j} \le \maxGroupSize&\forall \period{u} \in \periodSet~\forall\group{u}{j}\in\groupPeriodSet{u}\label{groupSizeDefined2}\\
    & \sum_{k \in \binPeriodSet{u}} y_{ujk} = 1 &\forall \period{u} \in \periodSet~\forall\group{u}{j}\in\groupPeriodSet{u}\label{groupIsScheduled}\\
    & c_{ujk} \le \groupSize{u}{j}& \forall \period{u} \in \periodSet~\forall\group{u}{j}\in\groupPeriodSet{u}~\forall k \in \binPeriodSet{u}\label{DefineContribution1}\\
    & c_{ujk} \le \period{0} \cdot y_{ujk} & \forall \period{u} \in \periodSet~\forall\group{u}{j}\in\groupPeriodSet{u}~\forall k \in \binPeriodSet{u} \label{DefineContribution2}\\
    & c_{ujk} \ge \groupSize{u}{j} - \period{0} \cdot (1 - y_{ujk}) & \forall \period{u} \in \periodSet~\forall\group{u}{j}\in\groupPeriodSet{u}~\forall k \in \binPeriodSet{u} \label{DefineContribution3}\\
    & p_{uk} = \sum_{\forall\group{u}{j}\in\groupPeriodSet{u}} c_{ujk} & \forall \period{u} \in \periodSet~\forall k \in \binPeriodSet{u} \label{PeriodLoadDefined}\\
    & \sum_{\forall \period{u} \in \periodSet} p_{u\left[k\mod \binPeriodSize{u}\right]} \le \cmax & \forall k \in \binPeriodSet{r-1} \\
    & \cmax \in \mathbb{Z} \label{obj2}\\
    & x_{uij} \in \binary& \forall \period{u} \in \periodSet~\forall\task{i}\in\taskPeriodSet{u}~\forall \group{u}{j}\in\groupPeriodSet{u} \\
    & z_{uj}\in\binary,~\groupSize{u}{j} \in \integers &\forall \period{u} \in \periodSet~\forall\group{u}{j}\in\groupPeriodSet{u}\\
    & y_{ujk}\in \binary,~c_{ujk}\in\integers  & \forall \period{u} \in \periodSet~\forall\group{u}{j}\in\groupPeriodSet{u}~\forall k \in \binPeriodSet{u}\\
    & p_{uk} \in \integers & \forall \period{u} \in \periodSet~\forall k \in \binPeriodSet{u}
\end{align}
\endgroup
}

 Each group is scheduled in exactly one suitable observation interval via $y_{ujk}$, see \cref{groupIsScheduled}. Even empty groups are scheduled, but with $\group{u}{j}=0$, they do not affect the solution. The contribution of group $\group{u}{j}$ to utilization of the observation interval $k$ is tracked via $c_{ujk}$ variables. Due to the dynamic group size, this is done with a big-M approach in \cref{DefineContribution1,DefineContribution2,DefineContribution3}; $c_{ujk}=\groupSize{u}{j}$ if $y_{ujk}1$, and zero otherwise. Finally, individual contributions are combined to obtain the load $p_{uk}$ of the given observation interval by groups with period $\period{u}$ in \cref{PeriodLoadDefined}, which, summed across periods, forms $\cmax$ \eqref{obj2}.

Note that a simpler model, based on \citeasnoun{lukasiewycz2009flexray},
\citeasnoun{grus2024icores} can be utilized for two special cases: when each task is in an individual group (obtaining the upper bound on $\cmax$). Only one group is necessary for any number of tasks with period $\period{u}$ in the observation interval $k$  (obtaining the lower bound on $\cmax$).

\section{Experiments}

We performed experiments with several synthetic instances, derived from those in \citeasnoun{grus2024icores}. We varied parameter values of $\header$ and $\maxGroupSize$, and we obtained 47 example instances with 50 - 600 tasks and up to six distinct periods. We compared three solvers: Gurobi v10.0, CP-SAT v9.14 of OR-Tools, and CP Optimizer v22.1. We were mostly interested in comparing MILP and CP solvers, since in our previous work \cite{grus2024icores} on PSP without grouping, CP Optimizer was slightly better than Gurobi.

\vspace{-5pt}
\begin{table}
\small
 \centering
\begin{tabular}{l|c|c|c|c}\toprule
& \# successes & mean $bg$ & median $bg$ & mean rank \\
\midrule
Gurobi & 47 & 0.10 & 0.00 & 1.15 \\
CP-SAT & 42 & 6.13 & 1.17 & 2.13 \\
CP Optimizer & 40 & 1.90 & 0.00 & 1.72 \\
\bottomrule
\end{tabular}
 \caption{Performance of the solvers for 47 instances. Best gap $bg \left[\%\right]$ is gap between method's $\cmax$ and best $\cmax$. Rank is calculated by sorting results for each instance.}
 \label{tab:experiments}
\end{table}
\vspace{-20pt}
As is shown in \cref{tab:experiments}, the Gurobi MILP solver was able to slightly outperform the CP-SAT solver of OR-Tools and CP Optimizer (both utilizing the same MILP-like model). This is true both for the averaged best gap $bg$ and the averaged rank of the methods.

\begin{figure}
    \centering
    \begin{subfigure}[b]{0.48\textwidth}
        \centering
        \includegraphics[width=0.78\textwidth]{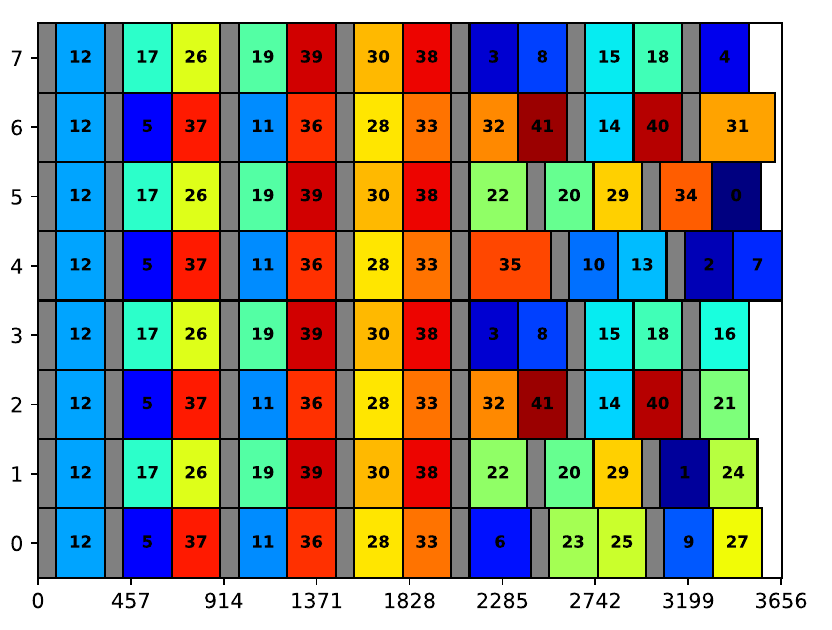}
        \caption{Example solution. Header size was set to 90 and maximum group size to 600. Headers are shown as gray tasks at the start of each (implicit) group.}
        \label{fig:example}
    \end{subfigure}
    \hfill
    \begin{subfigure}[b]{0.48\textwidth}
        \centering
        \includegraphics[width=0.78\textwidth]{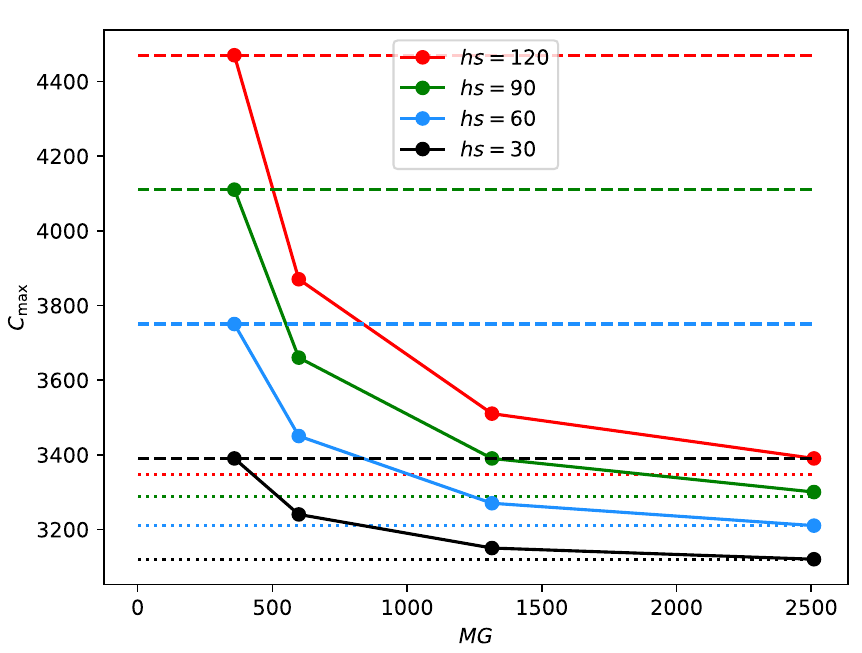}
        \caption{Parameter sensitivity analysis showing effect of $\header$ and $\maxGroupSize$ on $\cmax$. Dashed lines are solutions obtained for the upper-bound model, and dotted lines for the lower-bound model.}
        \label{fig:perf}
    \end{subfigure}

    \caption{Example solution and parameter sensitivity analysis for Gurobi solver on instance with $\periodSet=\left\{4000\cdot i~|~\forall i \in \left\{1,2,4,8\right\}\right\}$. Each solution was optimized for 5 minutes.}
    \label{fig:exp}
\end{figure}

In \cref{fig:exp}, several results of the Gurobi MILP solver are presented. In \cref{fig:example}, we can see how the maximum group size and headers affect the schedule. The header of size 90 is required for groups with a maximum size of 600. Headers are shown as gray rectangles at the start of the block of tasks implicitly included in the group.

In \cref{fig:perf} we can see how the increase of the maximum group size $\maxGroupSize$ improves the objective $\cmax$. Furthermore, the effect of different header sizes $\header$ is quite predictable, essentially moving the curves upwards with steeper gradients. Finally, we can see the upper and lower bound solutions with dashed and dotted lines, respectively. 

\section{Conclusion}
It is clear that grouping of signals offers significant advantages in real-life communication settings by improving resource utilization. We solved the problem using MILP and CP solvers. In the future, we will investigate the multi-resource setting \cite{grus2025} and the usage of different values of $\maxGroupSize$ per period. Furthermore, we will study special cases, including instances with divisible group sizes and periods for which the second-level problem is polynomial.

\section*{Acknowledgments}
This work was supported by the Grant Agency of the Czech Technical University in Prague, grant No.~SGS25/144/OHK3/3T/13 and co-funded by the European Union under the project ROBOPROX (reg. no. CZ.02.01.01/00/22\_008/0004590). 

\bibliographystyle{agsm}
\bibliography{biblio.bib}

\end{document}